\documentclass{aa}
\usepackage{graphics,txfonts}

\newcommand{\te}{$T_{\rm eff}$}
\newcommand{\tev}[1]{$T_{\rm eff}= #1$~K}
\newcommand{\lgg}{$\log g$}
\newcommand{\lggv}[1]{$\log g= #1$}
\newcommand{\vs}{$v_{\rm e}\sin\,i$}
\newcommand{\hd}{HD\,133792}
\newcommand{\vip}{{\tt VIP}}
\newcommand{\kms}{km\,s$^{-1}$}
\newcommand{\bs}{$\langle B \rangle$}
\newcommand{\bz}{$\langle B_{\rm z} \rangle$}
\newcommand{\ei}{E$_{\rm i}$}
\newcommand{\dfrac}[2]{\frac{\displaystyle #1}{\displaystyle #2}}

\newcommand{\fifps}[2]{\centering\resizebox{#1}{!}{\includegraphics{#2}}}
\newcommand{\figps}[1]{\resizebox{\hsize}{!}{\rotatebox{0}{\includegraphics{#1}}}}

\newcommand{\beq}{\begin{equation}}
\newcommand{\eeq}{\end{equation}}

\begin{document}

\title{Chemical stratification in the atmosphere of Ap star \hd%
\thanks{Based on observations collected at the European Southern Observatory,
Paranal, Chile (ESO programme No. 68.D-0254)}}
\subtitle{Regularized solution of the vertical inversion problem}

\author{O. Kochukhov\inst{1} \and V. Tsymbal\inst{2,3} \and 
        T. Ryabchikova\inst{4,3} \and V. Makaganyk\inst{2} \and S. Bagnulo\inst{5}}
 
\offprints{O. Kochukhov, \email{oleg@astro.uu.se}}

\institute{Department of Astronomy and Space Physics, Uppsala University, SE-751 20, Uppsala, Sweden
      \and Tavrian National University, Yaltinskaya 4, 95007 Simferopol, Crimea, Ukraine
      \and Department of Astronomy, University of Vienna, T\"urkenschanzstra{\ss}e 17, 1180 Vienna, Austria
      \and Institute of Astronomy, Russian Academy of Sciences, Pyatnitskaya 48, 109017 Moscow, Russia
      \and European Southern Observatory, Casilla 19001, Santiago 19, Chile}

\date{Received 15 May 2006 / Accepted 27 August 2006}

\abstract
{
High spectral resolution studies of cool Ap stars reveal conspicuous anomalies of the shape
and  strength of many absorption lines. This is a signature of large atmospheric chemical gradients
(chemical stratification) produced by the selective radiative levitation and gravitational settling of chemical species. 
}
{
Previous observational
studies of the chemical stratification in Ap stars were limited to fitting simple parametrized
chemical profiles. Here we present a new approach to mapping the vertical chemical structures
in stellar atmospheres. 
}
{
We have developed a regularized chemical inversion procedure that
uses all information available in high-resolution stellar spectra. The new technique for the
first time allowed us to recover chemical profiles without making \textit{a priori}
assumptions about the shape of chemical distributions. 
We have derived average abundances and applied the vertical inversion procedure to the 
high-resolution VLT UVES spectra of the weakly magnetic, cool Ap star \hd.
}
{
Our spectroscopic analysis yielded improved estimates of the atmospheric parameters of \hd.
We show that this star has negligible \vs\ and the mean magnetic field modulus 
\bs\,=\,$1.1\pm0.1$~kG. We have derived average abundances for 43
ions and obtained vertical distributions of Ca, Si, Mg, Fe, Cr, and Sr. 
All these elements except Mg show high overabundance in the deep layers and 
solar or sub-solar composition in the upper atmosphere of \hd. In contrast, the Mg abundance 
increases with height. 
}
{
We find that transition from the
metal-enhanced to metal-depleted zones typically occurs in a rather narrow range of depths in
the atmosphere of \hd. Based on the derived photospheric abundances, we conclude that \hd\ 
belongs to the rare group of evolved cool Ap stars, which possesses very large Fe-peak enhancement, but
lacks a prominent overabundance of the rare-earth elements.
}


\maketitle

\section{Introduction}
\label{intro}

Michaud (\cite{M70}) considered in detail a process of element separation in stellar
atmospheres under the mutual action of radiative acceleration and gravitational settling. He showed that in 
atmospheres with suppressed convection and turbulence this process could lead to chemical anomalies in the form
of under- or overabundances. This \textit{radiative diffusion} hypothesis was first applied to interpret the 
abundance anomalies inferred for the atmospheres of magnetic chemically peculiar Ap stars (Michaud et al.
\cite{MRC74}), which were known to be slow rotators and whose atmospheres are stabilized by the strong magnetic field. 

Diffusion processes are not only responsible for the observed average atmospheric abundance anomalies but also lead
to an inhomogeneous abundance distribution through the stellar atmosphere depending on the balance between radiative
acceleration and surface gravity. Borsenberger et al. (\cite{BPM81}) calculated Ca and Sr abundance profiles for
atmospheres with \te$\ge$10000~K and demonstrated the influence of chemical stratification on the profile of the
resonance  \ion{Ca}{ii} 3933~\AA\ line. These interesting theoretical predictions notwithstanding, the absence of
high-resolution, high signal-to-noise spectroscopic observations at that time did not permit a direct comparison between
observations and diffusion calculations. Later Babel (\cite{Babel92}) carried out detailed diffusion calculations
for Ca, Ti, Cr, Mn, and Sr in \te=8500~K stellar atmosphere and applied his model to explain an unusual shape of the
\ion{Ca}{ii} 3933~\AA\ line, which often shows extremely wide wings and a very sharp narrow core in the spectra of Ap
stars. Babel's calculations inspired the use of a simple approximation of the vertical element distribution in the
stellar atmosphere in the form of a step function. This parametrization is now widely employed in observational
stratification studies, for example Mn in HgMn stars (Sigut \cite{sigut01}), Si, Ca, Cr, Fe in Ap stars (Wade et
al. \cite{WLRK03}; Ryabchikova et al. \cite{RPK02,RLK05}). 

Despite the encouraging success of the previous attempts to model chemical gradients using the prescribed parametrized shape,
real element distributions in the atmospheres of peculiar stars may differ considerably from the simple step function, at least for
some chemical elements. Some theoretical studies also hinted at the possibility of complex vertical abundance
distributions (e.g. fig.~6 in Borsenberger et al. \cite{BPM81}). Moreover, recent self-consistent model atmosphere
calculations including diffusion (LeBlanc \& Monin \cite{LM04}) have identified a number of additional effects, like
mass loss and weak turbulent mixing, which may significantly affect the shape of the vertical chemical profiles but are
difficult to predict \textit{ab initio}. These interesting hydrodynamical effects, treated as free parameters in the current
theoretical diffusion modelling, could be potentially constrained by comparison with the observed vertical abundance
distributions.

Meanwhile modern spectrographs at large telescopes, such UVES at the ESO VLT, have reached remarkable precision and
spectral coverage, and now allow us to record very high signal-to-noise ratio data covering the whole optical region spectra
of moderately bright Ap stars. This dramatic improvement in the quality and quantity of the available observational
material suggests that unprecedented details of  the stellar atmospheric properties, and, in particular, the vertical
abundance distributions, can be deduced with the help of detailed modelling of the spectral line profiles.
These facts stimulated a recent surge of interest in the observational analysis of chemical stratification and emphasized
the need to improve stratification modelling techniques. 

In this paper we present a new procedure to study vertical chemical inhomogeneities in Ap-star atmospheres. 
We have chosen \hd\ (HR~5623, HIP~74181) for our modelling. This star was classified as an A0p Sr-Cr-Eu
object by Jaschek \& Jaschek (\cite{JJ59}). Mathys (\cite{M90}) noted that \hd\ exhibits remarkably sharp lines 
and no evidence of a strong magnetic field. Martinez \& Kurtz (\cite{MK94}) detected no rapid pulsational 
variability in \hd. Despite its brightness and extremely low \vs, facilitating detailed spectroscopic studies,
the star was never analysed with the model atmosphere method.

The paper is structured as follows. The principles and numerical details of the vertical inversion procedure are
given in Sect.~\ref{vip}. Observations of \hd\ and spectra reduction are described in Sect.~\ref{observ}.
Fundamental stellar parameters are determined in Sect.~\ref{params} and the abundance analysis is presented in
Sect.~\ref{abund}. We derive the vertical distribution of chemical elements in Sect.~\ref{strat}. Sect.~\ref{discus}
concludes the paper with the summary and discussion.

\begin{figure}[!th]
\figps{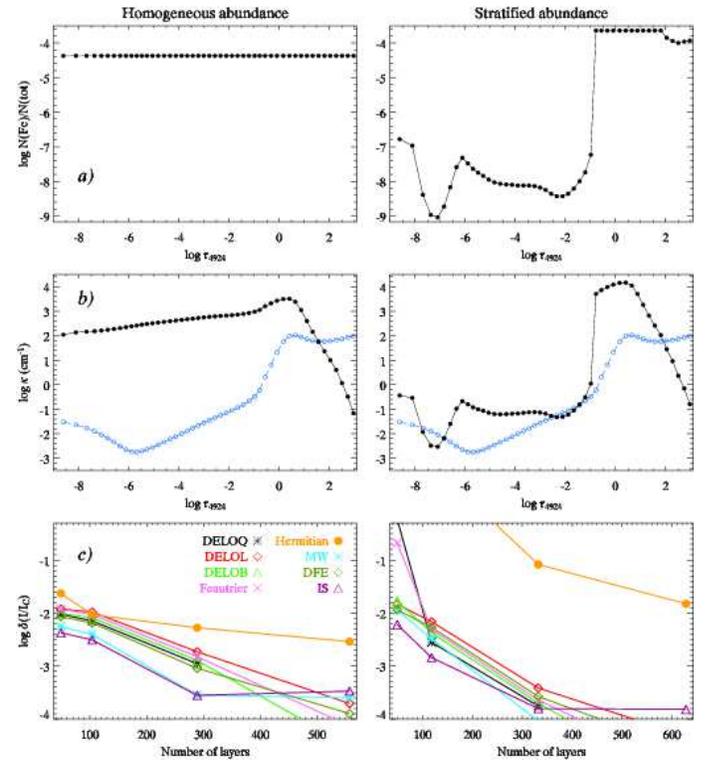}
\caption{Depth variation of the iron abundance and opacities for a chemically homogeneous
(\textit{left column}) and stratified (\textit{right column}) stellar atmosphere. {\bf a)} Fe
abundance as a function of the optical depth in continuum at $\lambda$ 4924~\AA; {\bf b)} depth
dependence of the line centre (\textit{filled symbols}) and continuous (\textit{open symbols}) 
opacities. The bottom panels ({\bf c}) show the maximum error in the normalized intensity profile 
of the \ion{Fe}{ii} 4923.93~\AA\ line computed with different radiative transfer algorithms and different 
number of layers in the model atmosphere.}
\label{fig1}
\end{figure}

\begin{figure*}[!th]
\figps{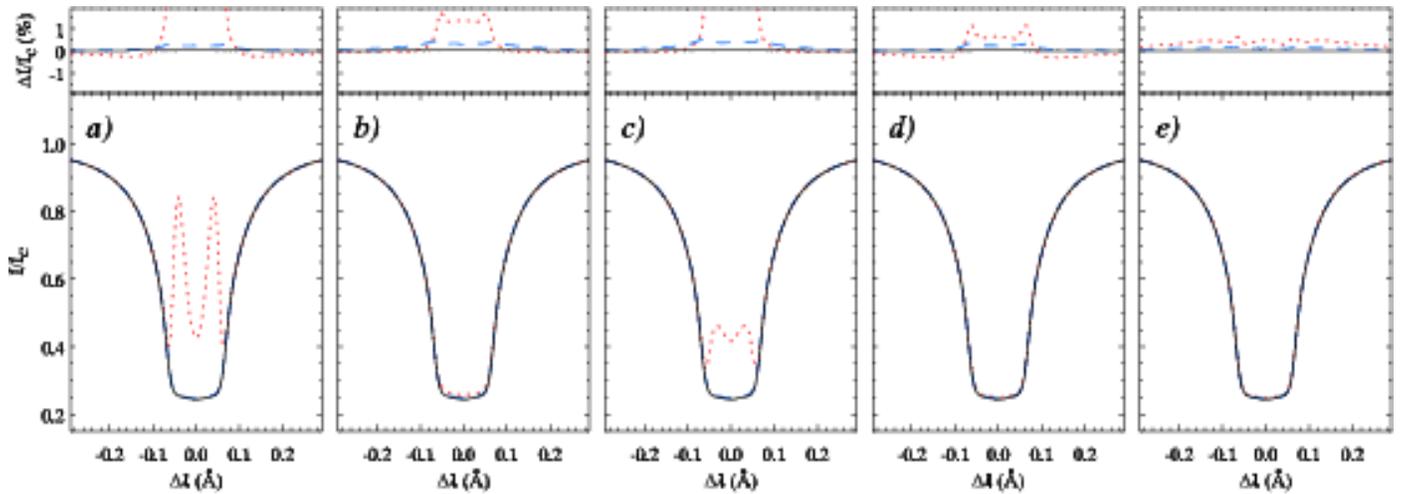}
\caption{Normalized disk-centre intensity profile of the \ion{Fe}{ii} 4923.93~\AA\ line computed for the stratified iron
abundance distribution shown in Fig.~\ref{fig1} using different radiative transfer algorithms: {\bf a)} DELOQ,
{\bf b)} DELOB, {\bf c)} Feautrier, {\bf d)} MW, {\bf e)} IS. In each plot the lower panel shows the line 
profiles computed for the model atmosphere with 49 (\textit{dotted line}), 118 (\textit{dashed line}), and
332 (\textit{solid line}) layers. The upper panels show the difference with respect to the computation
for the model atmosphere discretized into 628 layers.}
\label{fig2}
\end{figure*}

\section{The vertical inversion procedure}
\label{vip}
 
\subsection{Radiative transfer in a chemically stratified atmosphere}

The stability of the atmosphere of magnetic stars facilitates efficient radiative
diffusion and may lead to a buildup of significant chemical abundance gradients in the line-forming
regions. Theoretical calculations (Babel \cite{Babel92}; LeBlanc \& Monin \cite{LM04}) and
interpretation of the observations of cool Ap stars (e.g., Ryabchikova et al. \cite{RPK02,RLK05})
point to the presence of steep photospheric abundance gradients: variation by up to 2--3 orders of
magnitude may occur for some elements over a small range of optical depths. This extreme vertical
chemical non-uniformity has to be carefully accounted for in the solution of the radiative transfer
equation. It is not clear to what extent existing numerical radiative transfer schemes are able to
cope with a strong depth dependence of the line absorption coefficient and to generate sufficiently accurate
theoretical spectra. Since most of the methods rely on discretizing the stellar atmosphere into a number
of layers, performance of the numerical radiative transfer techniques is expected to improve with the
number of layers. In the context of the vertical inverse problem solution attempted here, the best
radiative transfer algorithm is the one achieving acceptable accuracy with the least number of vertical
zones in the chemically stratified model atmosphere.

We have evaluated the performance of 8 different radiative transfer algorithms commonly used in simulating
stellar spectra: the DELO solution with linear (DELOL, Rees et al. \cite{RMD89}), quadratic (DELOQ,
Socas-Navarro et al. \cite{STR00}) and Bezier spline (DELOB, Piskunov, private communication)
interpolation formulae for the source function; the Feautrier method (Rees et al. \cite{RMD89}); Hermitian
solution (Ruis-Cobo et al. \cite{RBC99}); stepwise Unno solution (MW, Martin \& Wickramasinghe
\cite{MW79}); Discontinuous Finite Element method (DFE, Castor et al. \cite{CDK92}) and the integral
solution of the radiative transfer equation as implemented in the {\tt ATLAS9} model  atmosphere code
(IS, Kurucz \cite{K93}). The disk-centre intensity across the \ion{Fe}{ii} 4923.93~\AA\  line was
calculated with each of the algorithms assuming an LTE source function and using an identical set of 
line and continuous opacities. The latter were calculated using the {\tt SYNTH3} code 
(Kochukhov \cite{K07}; Ryabchikova et al. \cite{RLK05}) for the 49-layer \tev{7700}, \lggv{4.0}
chemically stratified model atmosphere kindly provided by LeBlanc \& Monin (\cite{LM04}). In
this theoretical model the Fe distribution changes from $\log(Fe/N_{\rm tot})=-3.6$ at the bottom of
the atmosphere to $\log(Fe/N_{\rm tot})=-9.0$ in the uppermost layers with a large gradient at
$\log\tau_{4924}\approx-1.0$. For comparison we also evaluated the disk-centre intensity with the
same model atmosphere and homogeneous Fe abundance $\log(Fe/N_{\rm tot})=-4.4$. The depth dependence of
the iron concentration and of the line and continuous absorption coefficients are illustrated in
Fig.~\ref{fig1}a and b, respectively.

Calculation of the \ion{Fe}{ii} 4923.93~\AA\ line profile was carried out with the original 49-layer
depth grid and for the models interpolated onto a finer grid, containing 118, 332, and 628 layers. 
Interpolation employed the adaptive depth grid refinement procedure of the {\tt SYNTH3} code,
which subdivides layers where the line
absorption coefficient changes rapidly. The maximum error of each of the resulting \ion{Fe}{ii} line profiles
was recorded relative to the DELOQ solution for the 628-layer grid. The \ion{Fe}{ii} line shapes
corresponding to the stratified case are presented in Fig~\ref{fig2}. As expected, little discrepancy
is found between the spectra computed for the model with the maximum number of grid points. This
verifies the internal consistency of the radiative transfer schemes. However, as illustrated in
Fig.~\ref{fig1}c, the algorithms show rather different convergence properties. In the case of the
chemically homogeneous atmosphere, most of the methods already provide an accuracy of 1\,\% or better for
the original 49-layer model. At the same time, the stratified case turns out to be considerably more
challenging: all methods except Kurucz's integral solution require at least 100 layers in the model
atmosphere to achieve the necessary accuracy. On the other hand, the former method attains an acceptable 
maximum error of $\approx0.5$\,\% for the original 49-layer model atmosphere.

The outcome of our numerical tests highlights the stability and efficiency of the IS radiative transfer
algorithm. In principle, each of the studied numerical schemes can be successfully used for the
\textit{forward modelling} of the effects of chemical stratification if a sufficiently fine vertical
discretization is used. The advantage of the IS method is to provide the required accuracy at the least
cost. This is why we consider this algorithm to be the optimum choice for the \textit{vertical inversion
procedure} (\vip) which reconstructs a non-parametrized chemical distribution using a fixed depth grid. Taking
these results into account, we have adopted the IS algorithm in our vertical inversion code described
below.

\subsection{Spectrum synthesis}

Synthetic spectrum calculation for a chemically inhomogeneous stellar atmosphere is based on the {\tt
SYNTHV} code written by Tsymbal (\cite{T96}). The code accepts an LTE plane-parallel 
model atmosphere in the {\tt ATLAS9} format. In addition, the  input of \vip\
consists of a configuration file that specifies wavelength regions of interest and indicates the
respective line lists. Atomic line parameters are extracted from the VALD database (Kupka et al.
\cite{VALD2}). 

The code calculates new number densities for a given abundance stratification of any number of chemical
elements. Partition functions are calculated using the tables given in the PFSAHA subroutine of the {\tt
ATLAS9} code (Kurucz \cite{K93}). For some of the rare-earth elements we have updated these data
following Cowley \& Barisciano (\cite{CB94}). For calculations of the ionic populations we take into
account up to 6 ionization states of light elements (atomic number $\leq 28$) and up to 4 states for
heavy  elements. Calculation of the continuum opacity is performed for the blue and red edges of each of
the wavelength intervals and is based on the {\tt ATLAS9} subroutine KAPP (Kurucz \cite{K93}). We include
\ion{H}{i}, \ion{He}{i}, \ion{He}{ii}, H$^-$, different metallic bound-free and free-free transitions,
Rayleigh scattering for \ion{H}{i}, \ion{He}{i} and H$_{2}^{+}$, and electron scattering.

For each wavelength region the \vip\ code calculates the line opacity due to all atomic lines included in
the input line lists. Radiative damping, Stark broadening and van der Waals broadening are taken
into account. For all lines, except the \ion{He}{i} and hydrogen lines, we use the Voigt profile to
approximate the line opacity and adopt the broadening constants supplied by VALD or calculated
according to the classical expressions (see Gray \cite{G92}). The Stark broadening of the hydrogen
lines is  estimated by interpolating in the tables by Lemke (\cite{L97}). For the \ion{He}{i} we use
the Stark broadening profiles of Barnard et al. (\cite{BCS74}) when available.

The radiative transfer equation is solved with the modified {\tt ATLAS9} subroutine JOSH.
The specific intensity is calculated for 7 angles between the local outward normal and the line of sight. Each
intensity profile is convolved with appropriate functions to model stellar rotation,
radial-tangential  macroturbulent broadening, and the instrumental profile. The flux spectrum is
produced by integrating specific intensities over the stellar disk. 

Different spectral intervals are treated independently, which permits us to use different instrumental
broadening profiles and radial velocity corrections.

\subsection{Inverse problem solution}

We must solve an inverse problem in order to find the distribution of a chemical element with depth in the stellar 
atmosphere. We seek the vertical chemical profile that would provide an adequate description of the
spectral line intensities and profiles of all spectral features belonging to a given species. This problem
is similar to abundance Doppler imaging (Kochukhov et al. \cite{KDP04}), except that the aim is to
obtain the best-fit chemical stratification instead of a 2-D picture of the horizontal abundance
distribution. 

Similar to Doppler imaging, the vertical inversion belongs to the class of ill-posed problems
(Tikhonov \& Arsenin \cite{TA77}). A given sample of spectral lines recorded in the optical spectra of
A-type stars is characterized by a very uneven sensitivity to the properties of different atmospheric
layers. Typically, the chemical composition of the line forming regions  (optical depth
$-2\le\log\tau_{5000}\le0$) is very well constrained, whereas very few or no lines sample the uppermost
and deepest layers. This presents a formidable difficulty for a vertical inversion algorithm: a multitude of
widely different, high-contrast solutions can provide a reasonable description of observations.
Furthermore, inversions are unstable with respect to the initial guess, and the properties of
solutions, especially the amplitude of abundance gradients, will strongly depend on the
vertical sampling of the stellar model atmosphere adopted in the inversion. 

A common procedure to alleviate this non-uniqueness and instability of the ill-posed problem is to
introduce a \textit{regularization}. This brings in additional \textit{a priori} information that
can be used to constrain the problem and decouple solution from the vertical sampling used in the model
atmosphere. For instance, one can look for a solution that maximizes the entropy (Vogt et al.
\cite{VPH87}) or impose a requirement that the solution must be a smooth function (Goncharskij et al.
\cite{GSK77}). The second criterion, also known as the Tikhonov regularization procedure, is appropriate for
the vertical inversion problem and is implemented in our code.

Thus, the vertical chemical inversion reduces to the mathematical problem of finding a depth-dependent
chemical distribution $\varepsilon(x)$ that minimizes the function
\beq
{\cal F}=\sum_i W_i \sum_\lambda 
         \left(F^{\rm obs}_{i\lambda} - F^{\rm syn}_{i\lambda}(\varepsilon)\right)^2/\sigma^2(F^{\rm obs}_{i\lambda})
         +\Lambda\sum_j \left (\dfrac{d\varepsilon}{dx}\right)_j^2.
\label{func}
\eeq
Here $F^{\rm obs}_{i\lambda}$ and $\sigma(F^{\rm obs}_{i\lambda})$ are the observed stellar spectrum and
its uncertainty. $F^{\rm syn}_{i\lambda}(\varepsilon)$ is the theoretical spectrum synthesis
for the chemical distribution $\varepsilon(x)\equiv\log(N/N_{\rm tot})$. 
We use the logarithm of the continuum optical depth at $\lambda=5000$~\AA\ as an
independent vertical variable $x\equiv\log \tau_{5000}$. The index $i$ runs over all spectral intervals,
and $\lambda$ -- over the wavelength points in each interval. We assign different weights $W_i$ to
spectral regions according to the quality of the respective observational data, or if there is a necessity
to emphasise or diminish the relative importance of fits to the specific spectral features. The
second term in Eq.~(\ref{func}) represents the Tikhonov regularization, and $\Lambda$ denotes the
regularization parameter. 

The chemical distribution $\varepsilon(x)$ is a discrete function that defines the concentration 
of a given element in each model atmosphere layer. Its shape is not prescribed in advance
but is constrained entirely by the available observations and by the Tikhonov regularization function.
The role of regularization is to ensure the stability of the inversion procedure and to provide the simplest
(smoothest) solution that fits observations.

\begin{figure}[!th]
\figps{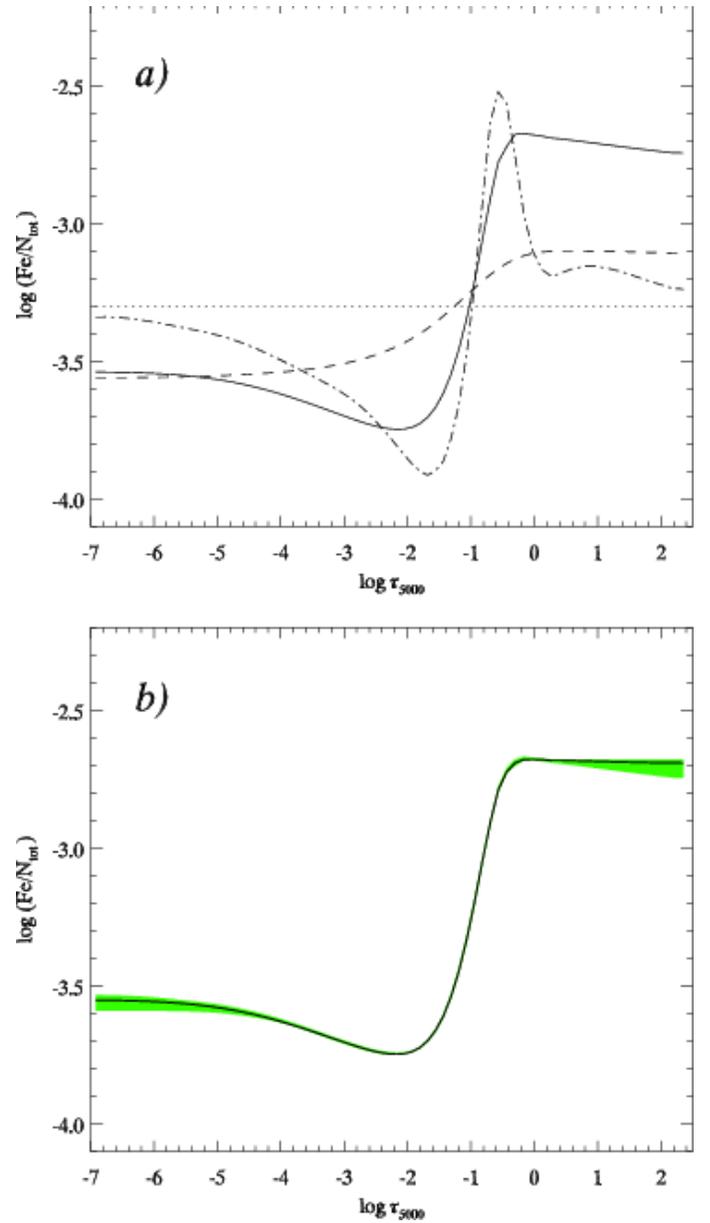}
\caption{{\bf a)} Dependence of the inferred iron vertical distribution on the
regularization parameter adopted in \vip\ reconstruction. \textit{Solid line:} reconstruction with the optimal regularization, 
\textit{dashed line:} inversion with 20 times higher regularization; \textit{dashed-dotted line:}
distribution obtained using 10 times smaller regularization. 
The horizontal dotted line shows homogeneous Fe distribution adopted as the initial guess.
{\bf b)} Reconstruction of the Fe distribution using optimal regularization and different homogeneous
initial guesses ($\log (Fe/N_{\rm tot})$ between $-2.5$ and $-4.0$). The solid curve shows the average
reconstructed iron stratification. The shaded area represents the full range of Fe abundance
for the inversions started from different initial values.}
\label{fig3}
\end{figure}

We use the Levenberg-Marquardt method (Press et al. \cite{numres}) as the minimization algorithm. It
combines the best features of the gradient search with the method of linearizing the fitting function,
which ensures rapid convergence close to the minimum. For the \vip\ code we have adapted the 
Levenberg-Marquardt routine described by Piskunov \& Kochukhov (\cite{PK02}). Trial inversions
demonstrate that \vip\ converges to a stable solution in no more than 10--15 iterations.

Derivatives of the fitting function with respect to the abundance in each layer, 
$\partial{\cal F}/\partial \varepsilon_j$, are required by the Levenberg-Marquardt algorithm and
are evaluated analytically for the regularization function. A numerical approximation is used for 
the first derivatives of the synthetic line profiles:
\beq
\dfrac{\partial F^{\rm syn}_{i\lambda}}{\partial \varepsilon_j} \simeq 
\dfrac{F^{\rm syn}_{i\lambda}(\varepsilon_j+\Delta\varepsilon)-F^{\rm syn}_{i\lambda}(\varepsilon_j)}{\Delta\varepsilon},
\eeq
where we use $\Delta\varepsilon=0.01$~dex.

The regularization parameter $\Lambda$ is determined empirically. We adjust regularization in such a way
that i) the solution $\varepsilon(x)$ is smooth but, at the same time, gives a satisfactory description 
of the stellar observations and ii) the solution is independent of the initial guess. In practice,
these requirements are fulfilled when the contribution of regularization to the total discrepancy function
${\cal F}$ is comparable to, but not larger than the first term of Eq.~(\ref{func}).

Figure~\ref{fig3} illustrates dependence of the vertical Fe abundance inversion in \hd\ on the choice of
regularization strength. The vertical distributions recovered with widely  different $\Lambda$ are shown
in Fig.~\ref{fig3}a. The solid curve corresponds to the results obtained with the optimum regularization.
Regularization that is too weak leads to large abundance gradients (dashed-dotted curve) and cannot
ensure uniqueness of the solution. On the other hand, a much stronger regularization (dashed line)
smoothes the inferred vertical distribution of Fe and does not allow us to achieve a low $\chi^2$ in the
fit to observations.

The stability test of the vertical inversion using the optimum regularization is presented in
Fig.~\ref{fig3}b. The thick solid line shows the average of 16 Fe distributions recovered in the
inversions started from constant Fe abundance $\log (Fe/N_{\rm tot})$ in the range from
$-2.5$ to $-4.0$~dex. The dark area represents the full span of the individual vertical solutions. Its
maximum width is 0.07~dex, and the mean width is 0.03~dex. Thus, despite drastically different initial
guesses, regularization has ensured convergence to nearly the same Fe stratification.

In addition to chemical stratification, elements are often distributed inhomogeneously over
the surface of magnetic chemically peculiar stars (e.g., Kochukhov et al. \cite{KDP04}).  In
principle, it is possible to compute a synthetic spectrum taking into account a  non-uniform
distribution of elements over the stellar surface and chemical stratification at the same time.
However, such a 3-D abundance mapping procedure requires complete rotational phase coverage and can
only be applied to rapidly rotating Ap stars. On the other hand, \hd\ investigated in our paper is
an extremely slow rotator (see Sect.~\ref{params}) and lacks evidence of spectroscopic
variability. Therefore, no information can be derived about possible spotted distribution of
elements over the surface of \hd, and this effect is disregarded in the present
study.

\section{Observations and data reduction}
\label{observ}

High-resolution, high signal-to-noise ratio spectra of \hd\
were obtained with the UVES instrument of the ESO VLT on 26 February 2002
in the program 68.D-0254. The UVES spectrometer is
described by Dekker et al. (\cite{DOK00}). The observations were carried out
using both available dichroic modes. The detailed log of the
observations is given in Table~\ref{Table_UVES_Log}. In both the blue
arm and the red arm the slit width was set to 0.5$^{\prime\prime}$, for a spectral
resolution of about 80\,000. The slit was oriented along the
parallactic angle, in order to minimize losses due to atmospheric
dispersion.  Almost the full wavelength interval from 3030 to 10400~\AA\,
was observed, except for a few gaps, the largest of which are at
5760--5835~\AA\ and 8550--8650~\AA. In addition, there are several small
gaps, about 1\,nm each, due to the lack of overlap between the
\'{e}chelle orders in the 860U setting.

The UVES data have been reduced with the automatic pipeline described in
Ballester et al. (\cite{BMB00}). For all settings, science frames are
bias-subtracted and divided by the extracted flat-field, except for
the 860~nm setting, where the 2-D (pixel-to-pixel) flat-fielding is used,
in order to better correct for the fringing. Because of the high flux
of the spectra, we used the UVES pipeline \textit{average extraction}
method.

All spectra were normalized to the continuum with an
interactive procedure that employed either a low-degree polynomial
or a smoothing spline function.

\begin{table}
\caption{Log of UVES observations of \hd.}
\label{Table_UVES_Log}
\begin{tabular}{ccc}
\hline 
\hline
Date   &    UT     & Setting (nm) \\
\hline
2002-02-26 & 07:06:01 & 346 \\
2002-02-26 & 07:07:59 & 346 \\
2002-02-26 & 06:59:57 & 437 \\
2002-02-26 & 07:01:43 & 437 \\
2002-02-26 & 07:06:00 & 580 \\
2002-02-26 & 07:08:07 & 580 \\
2002-02-26 & 06:59:57 & 860 \\
2002-02-26 & 07:01:50 & 860 \\
\hline
\end{tabular}
\end{table}

\section{Atmospheric parameters, rotation, and magnetic field}
\label{params}

\begin{figure*}[!th]
\figps{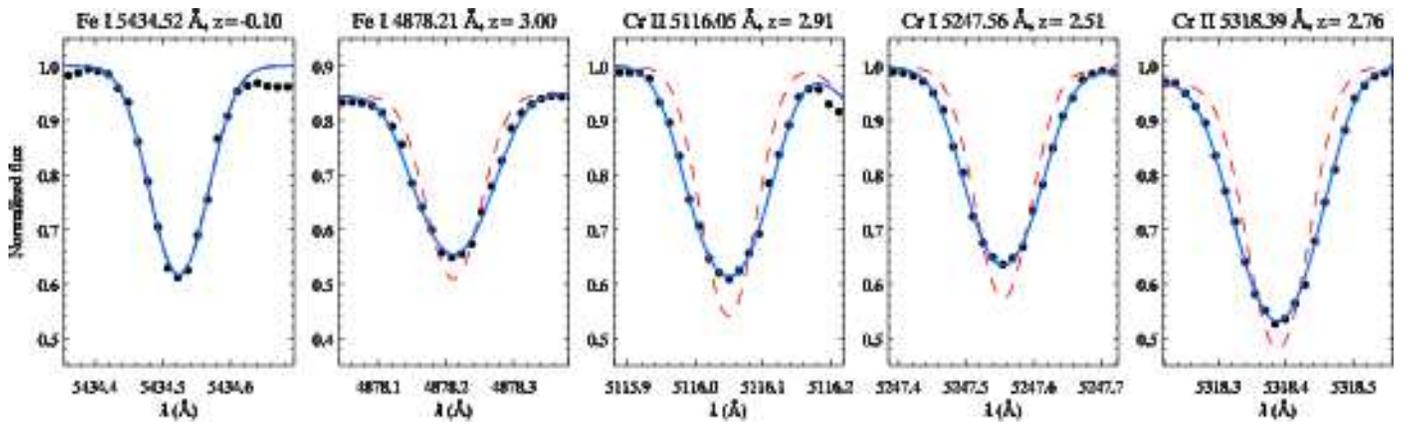}
\caption{Magnetic broadening of lines in the spectrum of \hd. In each panel observations
(\textit{symbols}) are compared with the theoretical spectra computed for \bs\,=\,0~kG
(\textit{dashed line}) and \bs\,=\,1.1~kG (\textit{solid line}). The leftmost panel shows
the \ion{Fe}{i} 5434.52~\AA\ line which has $z=-0.10$ and, therefore, is almost insensitive to the 
magnetic field effects. The other panels show \ion{Fe}{i}, \ion{Cr}{i}, and \ion{Cr}{ii} lines with 
large effective Land\'e factors.}
\label{fig4}
\end{figure*}

We used the Str\"omgren photometric indices of \hd\ to obtain an initial estimate of the stellar
model atmosphere parameters. The observed colours, $b-y=0.026$, $c_1=0.180$, $m_1=1.110$ (Hauck \&
Mermilliod \cite{HM98}), were dereddened adopting $E(B-V)=0.09$. This colour excess follows from
the reddening maps by Lucke (\cite{L78}) and  high-resolution dust maps by Schlegel et al.
(\cite{SFD98}). Taking into account H$\beta=2.866$  (Hauck \& Mermilliod \cite{HM98}), we have
established \tev{9334} and \lggv{3.84} with the calibration by Moon \& Dworetsky
(\cite{MD85}) implemented in the {\tt TEMPLOGG} code (Rogers \cite{R95}).  

Model atmospheres for \hd\ were calculated with the {\tt ATLAS9} code (Kurucz \cite{K93}),
using the ODF with 3  times the solar metallicity and zero microturbulent velocity. With 
these metal-enhanced models we further fine-tuned the stellar parameters to fit the
hydrogen H$\alpha$ and H$\beta$ lines. This procedure yielded the final parameters
\tev{9400\pm200} and \lggv{3.7\pm0.1}. The final model atmosphere of \hd\ has 72 layers and
covers optical depths from $\log\tau_{\rm 5000}=-6.9$ to $\log\tau_{\rm 5000}=2.4$.

The projected rotational velocity of \hd\ was established by synthetic spectrum fitting of the
magnetically insensitive \ion{Fe}{i} lines. In particular, for the \ion{Fe}{i} 5434.52~\AA\ line
(mean Land\'e factor $z=-0.01$) no rotational Doppler broadening appears to be necessary after the
instrumental smearing corresponding to  the resolution of our UVES observations is accounted for
(see Fig.~\ref{fig4}). The respective upper \vs\ limit is $\sim$\,1.0~\kms. 

Using the Hipparcos parallax of \hd\ ($\pi=5.87\pm0.66$~mas, Perryman et al. \cite{PLK97})
and adopting \te\ very close to the value established above, Kochukhov \& Bagnulo  (\cite{KB06})
determined $\lg L/L_{\odot}=2.02\pm0.10$ and $M=2.80\pm0.14\,M_{\odot}$. Comparison with the
theoretical stellar interior models (Schaller et al. \cite{SSM92}) suggests that \hd\ is
significantly evolved from the ZAMS and is likely to be close to the end of its main sequence
evolutionary phase. 

One derives $R=3.9\pm0.5\,R_{\odot}$ from the aforementioned value of the stellar luminosity and
\tev{9400\pm200}. This radius implies a lower limit of about 200~d for the rotation period if the
star is viewed equator-on.

No resolved Zeeman split spectral lines are found in the spectra of \hd, indicating a mean field strength of $\la$\,2~kG (Mathys
\cite{M90}). Mathys \& Lanz (\cite{ML92}) have also failed to detect relative intensification of the \ion{Fe}{ii} 6147 and
6149~\AA\ lines and, on this basis, have suggested a zero magnetic field for the star. However, detailed magnetic spectrum
synthesis calculations by Takeda (\cite{T91}) showed this interpretation to be invalid. For magnetic fields weaker than 
$\approx$\,2~kG, the diagnostic content of this particular \ion{Fe}{ii} line pair is highly ambiguous, and, in general, 
there is no correlation between the difference of the equivalent widths of the two \ion{Fe}{ii} lines and magnetic field
intensity.

With our high-quality spectra we see
clear indication of an extra broadening, invariably correlating with the magnetic sensitivity, for many spectral lines.
Examples of this are illustrated in Fig.~\ref{fig4} for several Fe and Cr lines with large Land\'e factors. We
have  carried out polarized radiative transfer calculations with the \mbox{\tt SYNTHMAG} code (Piskunov \cite{P99}) to
estimate the mean field strength. The best fit is achieved for \bs\,=\,$1.1\pm0.1$~kG, confirming our earlier field
strength measurement (Ryabchikova et al. \cite{RLK04}). 

Recently Kochukhov \& Bagnulo (\cite{KB06}) detected a
marginal positive longitudinal field of $\approx$\,120~G based on the two low-resolution spectropolarimetric observations
of \hd\ with FORS1 at VLT. The longitudinal field did not change sign nor appreciably vary in strength between the two
FORS1 \bz\ field measurements separated by 9 months. The large \bs\ to \bz\ ratio obtained for \hd\ hints
that currently this star is observed at the crossover phase.

\section{Average abundances of \hd}
\label{abund}

Our line identification is based on the theoretical spectrum calculated for the whole spectral region
3050--9000 \AA\  using the line extraction from VALD (Kupka et al. \cite{VALD2} and references
therein) and DREAM (Bi\'emont et al. \cite{DREAM}) databases. Atomic data on the REE elements
compiled in  the DREAM database were extracted via the VALD interface. Comparison of the
synthetic and  observed spectra allowed us to choose the least blended lines for the chemical abundance
and stratification analysis. 

Spectral lines in \hd\ are very sharp due to a negligible rotation and a rather small magnetic field 
intensity. Thus, for most elements we have measured equivalent widths and carried out abundance analysis  with
the Kurucz's {\tt WIDTH9} code, modified by one of us (VT) to accept the line lists in the
VALD output format. Only in the case of either blended  lines or lines affected by hyperfine and/or
isotopic splitting did we employ spectral synthesis for abundance determination (Be, \ion{Mn}{i},
\ion{Mn}{ii}, \ion{Eu}{ii}). 

The recent experimental oscillator strengths currently included in the VALD database were used for the
following elements: \ion{Ti}{ii} (Pickering et al. \cite{PTP01}), \ion{Mn}{ii} (Kling \& Griesmann
\cite{KG00};  Kling et al. \cite{KSG01}), \ion{Nd}{ii} (Den Hartog et al. \cite{HLSC04}), \ion{Dy}{ii}
(Wickliffe et al. \cite{WLN00}). For  \ion{Ce}{ii}, \ion{Ce}{iii}, \ion{Nd}{iii} oscillator strengths were
taken from the DREAM database (Palmeri et al. \cite{Ce2}, Bi\'emont et al. \cite{Ce3}, Zhang et al.
\cite{Nd3}), while for \ion{Pr}{iii} the calculations (Ryabtsev, private communication) based on the
extended energy levels analysis (Wyart et al. \cite{Pr3}) were used. For \ion{Eu}{ii} atomic data are
taken from Lawler et al. (\cite{Eu2}). Abundances from the \ion{Cr}{ii}, \ion{Fe}{ii}, \ion{Co}{ii} lines
were based on the oscillator strengths calculated with the orthogonal operator technique (Raassen \&
Uylings \cite{RU98}). Its advantage  was discussed by Ryabchikova et al. (\cite{RLK05}). 

Hyperfine splitting of the \ion{Mn}{i} (Blackwell-Whitehead et al. \cite{Mn1hfs05}) and \ion{Mn}{ii} (Holt et
al. \cite{HSR99}) lines was taken into account. However, its effect was found to be negligible for \ion{Mn}{i}
lines ($\le$0.04 dex) and very small for \ion{Mn}{ii} lines. For the abundance analysis we used
only \ion{Mn}{ii} lines for which experimental oscillator strengths and hfs constants are available. The
\ion{Co}{i} lines are too weak in the spectrum of \hd\ to be influenced by hfs.

\begin{table}[!th]
\caption{Atmospheric abundances in the Ap star HD\,133792 with the error
estimates based on $n$ measured lines. Abundances in the solar atmosphere and in the evolved Ap star HD\,204411
are given for comparison.}
\label{Abund}
\begin{footnotesize}
\begin{center}
\begin{tabular}{l|lr|l|c}
\noalign{\smallskip}
\hline
\hline
Ion &\multicolumn{2}{c|}{\hd} & HD\,204411  & Sun \\                                  
    &$\log (N/N_{\rm tot})$ & $n$ &$\log (N/N_{\rm tot})$ & $\log (N/N_{\rm tot})$ \\       
\hline
\ion{Be}{ii}  &  $-$11.80:         &  2 &         & ~$-$10.66~ \\                            
\ion{C}{i}    & ~~$-$4.68$\pm$0.07 &  2 & ~$-$4.37& ~~$-$3.65~ \\			    
\ion{N}{i}    & ~~$-$5.16$\pm$0.08 &  2 & ~$-$3.93& ~~$-$4.26~ \\			    
\ion{O}{i}    & ~~$-$4.23$\pm$0.42 &  5 & ~$-$4.03& ~~$-$3.38~ \\			    
\ion{Na}{i}   & ~~$-$5.35$\pm$0.22 &  2 & ~$-$5.28& ~~$-$5.87~ \\			     
\ion{Mg}{i}   & ~~$-$3.91$\pm$0.02 &  5 & ~$-$4.34& ~~$-$4.51~ \\			     
\ion{Mg}{ii}  & ~~$-$4.18$\pm$0.68 &  2 & ~$-$4.62& ~~$-$4.51~ \\			     
\ion{Al}{ii}  & ~~$-$6.03:	   &  1 & ~$-$5.85& ~~$-$5.67~ \\			     
\ion{Si}{i}   & ~~$-$3.69$\pm$0.21 &  3 & ~$-$4.13& ~~$-$4.53~ \\			     
\ion{Si}{ii}  & ~~$-$5.09$\pm$0.50 &  5 & ~$-$4.11& ~~$-$4.53~ \\			     
\ion{Ca}{i}   & ~~$-$5.31$\pm$0.20 &  2 & ~$-$5.17& ~~$-$5.73~ \\			     
\ion{Ca}{ii}  & ~~$-$7.36$\pm$0.83 &  3 & ~$-$4.67& ~~$-$5.73~ \\			     
\ion{Sc}{ii}  & ~~$-$9.50$\pm$0.38 &  2 & ~$-$9.52& ~~$-$8.99~ \\			     
\ion{Ti}{ii}  & ~~$-$6.88$\pm$0.18 & 50 & ~$-$6.49& ~~$-$7.14~ \\			     
\ion{V}{ii}   & ~~$-$8.14$\pm$0.17 &  8 &         & ~~$-$8.04~ \\			     
\ion{Cr}{i}   & ~~$-$3.79$\pm$0.20 &131 & ~$-$4.85& ~~$-$6.40~ \\			     
\ion{Cr}{ii}  & ~~$-$3.75$\pm$0.22 &320 & ~$-$4.70& ~~$-$6.40~ \\			     
\ion{Mn}{i}   & ~~$-$5.55$\pm$0.17 & 13 & ~$-$5.96& ~~$-$6.65~ \\			     
\ion{Mn}{ii}  & ~~$-$5.39$\pm$0.10 &  9 & ~$-$5.66& ~~$-$6.65~ \\			     
\ion{Fe}{i}   & ~~$-$3.31$\pm$0.20 &247 & ~$-$3.76& ~~$-$4.59~ \\			     
\ion{Fe}{ii}  & ~~$-$3.18$\pm$0.25 &347 & ~$-$3.52& ~~$-$4.59~ \\			     
\ion{Co}{i}   & ~~$-$6.49$\pm$0.16 &  6 & ~$-$6.19& ~~$-$7.12~ \\			     
\ion{Co}{ii}  & ~~$-$5.99$\pm$0.79 &  6 & ~$-$6.50& ~~$-$7.12~ \\			     
\ion{Ni}{i}   & ~~$-$6.05$\pm$0.25 & 15 & ~$-$5.68& ~~$-$5.81~ \\			     
\ion{Ni}{ii}  & ~~$-$5.96$\pm$0.11 &  2 & ~$-$5.31& ~~$-$5.81~ \\			     
\ion{Sr}{i}   & ~~$-$5.71$\pm$0.37 &  8 & ~$-$7.74& ~~$-$9.12~ \\			     
\ion{Sr}{ii}  & ~~$-$6.36$\pm$0.09 &  3 & ~$-$8.5 & ~~$-$9.12~ \\			     
\ion{Y}{ii}   & ~~$-$9.12$\pm$0.46 &  6 & ~$-$9.95& ~~$-$9.83~ \\			     
\ion{Zr}{ii}  & ~~$-$9.47$\pm$0.10 &  7 & ~$-$8.66& ~~$-$9.45~ \\			     
\ion{Ru}{ii}  & ~~$-$8.82:	   &  1 &	   & ~$-$10.20~ \\			     
\ion{Pd}{i }  & ~~$-$7.33$\pm$0.10 &  2 &	   & ~$-$10.35~ \\			     
\ion{Ba}{ii}  & ~~$-$8.73$\pm$0.16 &  4 & ~$-$9.02& ~~$-$9.87~ \\			     
\ion{La}{iii} & ~~$\le-$9.6	   &  2 &	   & ~$-$10.46~ \\			      
\ion{Ce}{ii}  & ~~$-$9.07:	   &  1 & $-$10.26& ~$-$10.46~ \\			      
\ion{Ce}{iii} & ~~$-$8.41$\pm$0.07 &  3 &	  & ~$-$10.46~ \\			      
\ion{Pr}{iii} & ~~$-$9.51$\pm$0.17 &  3 & $<-$10.5& ~$-$11.33~ \\			      
\ion{Nd}{ii}  & ~~$-$9.08$\pm$0.43 &  3 & ~$-$9.48& ~$-$10.59~ \\			      
\ion{Nd}{iii} & ~~$-$9.10$\pm$0.01 &  2 & $-$10.05& ~$-$10.59~ \\			      
\ion{Sm}{ii}  & $\le-$10.4         &  1 &         & ~$-$11.03~ \\                            
\ion{Eu}{ii}  & $-$9.80$\pm$0.20   &  4 & $-$10.95& ~$-$11.52~ \\                             
\ion{Gd}{ii}  & ~~$-$9.60$\pm$0.04 &  2 &         & ~$-$10.92~ \\                             
\ion{Dy}{ii}  & $-$10.02$\pm$0.38  &  3 &         & ~$-$10.90~ \\			     %
\ion{Er}{ii}  & ~~$\le-$9.4        &  1 &         & ~$-$11.11~ \\			     %
\hline											     %
\te     &\multicolumn{2}{c|}{9400~K}   & 8400~K      & 5777~K  \\				
\lgg    &\multicolumn{2}{c|}{3.70~~~~~}& 3.50        & 4.44~~~~\\				   
\bs     &\multicolumn{2}{c|}{1.1~kG}   &$\le$0.7~kG  & 	       \\				
\hline											  
\end{tabular}
\end{center}
\end{footnotesize}
\end{table}

The final results of the abundance analysis assuming chemical homogeneity of the \hd\ atmosphere 
are presented in Table\,\ref{Abund}. We have compared them with the atmospheric abundances in
another evolved, slightly cooler Ap star with a very weak magnetic field, HD~204411 (Ryabchikova et al.
\cite{RLK05}), and in the solar atmosphere (Asplund et al. \cite{NSA05}).  

The light elements Be, C, N, and O, are underabundant by an order of magnitude in \hd\ when compared with the
chemical composition of the solar photosphere. The UVES spectra allowed us to investigate the resonance
\ion{Be}{ii} lines, which are only marginally visible in \hd, indicating an order of magnitude Be deficiency.
Thus, on the basis of much more accurate spectroscopic observations we have confirmed the earlier coarse assessment of
a Be underabundance in the atmospheres of 4 cool magnetic Ap stars (Gerbaldi et al. \cite{GFM86}), which contrasts with 
the very high Be overabundance derived for non-magnetic hot HgMn stars (Boesgaard et al. \cite{BHW82}). 
The upper limit for the He abundance, obtained from the \ion{He}{i}
$\lambda$4471 line, is $\log (He/N_{\rm tot})$\,=\,$-2.5$, which is 1.5~dex lower than the solar value.
The large underabundance of light elements is typical of cool Ap stars (Gerbaldi et al. \cite{O1-89}; Roby \&
Lambert \cite{RL90}). 

Abundances of Mg, Si and Ca derived from the lines of neutrals and first ions differ significantly, which
is an indication of chemical stratification (see Ryabchikova et al. \cite{RWL03}). 

The Fe-peak elemental abundances are similar in both \hd\ and HD~204411, with more pronounced Fe and, in
particular, Cr anomalies in \hd. This is in agreement with the general dependence of the Cr and Fe abundance on
the effective temperature in Ap stars (Ryabchikova et al. \cite{abun04}; Ryabchikova \cite{abun05}). We did
not notice any significant dependence of the individual line abundances on the wavelength or excitation
energy for Cr lines. On the other hand, such a dependence exists for Fe lines: abundances derived from the low
excitation lines below the Balmer Jump (BJ) are systematically smaller than those derived from the lines above
the BJ. This is another signature of possible stratification of elements. 

Special attention was given to the careful study of the rare-earth elements. The evolved Ap star HD~204411
is very Cr-Fe overabundant and REE-poor compared to the majority of magnetic Ap stars, where a large REE
overabundance is typical. REE lines are very weak in the spectrum of \hd\ and  reliable
abundances can be derived only for the lines of \ion{Ce}{iii}, \ion{Pr}{iii}, \ion{Nd}{ii}, \ion{Nd}{iii},
and \ion{Eu}{ii}. We find a REE overabundance of about 1.5~dex relative to the solar composition, with a
hint of decreasing overabundance towards the heavier species. 

\section{Stratification in \hd}
\label{strat}

For the stratification analysis we have chosen chemical elements with a large number of unblended spectral
lines (Cr and Fe) and elements showing the most conspicuous discrepancies between the line-by-line
abundances derived in the approximation of a chemically homogeneous atmosphere (Ca, Mg, Si, Sr). 

The lack of the accurate atomic data precludes a stratification analysis for some other elements showing
discrepant abundances. For instance, a significantly different abundance is derived from the \ion{Co}{ii}
lines with different excitation energies. However, hyperfine splitting strongly affects these lines, and
the hfs parameters are not known for many \ion{Co}{ii} transitions.

Stratification calculations were performed for 6 elements. Atomic parameters of the spectral lines employed
for each of the element are listed in Table\,\ref{strat-list}. The central wavelength, excitation potential, 
oscillator strength, and the Stark broadening constant are provided. For the \ion{Cr}{ii} and \ion{Sr}{ii}
lines marked by asterisks in Table\,\ref{strat-list}, experimental Stark broadening data were taken from
Rathnore et al. (\cite{Stark-Cr}) and Fleurier et al. (\cite{Stark-Sr}), respectively. For all other lines
the broadening constants were either extracted from the VALD database or calculated using the classical
approximation (Gray \cite{G92}).

\begin{table*}[!th]
\caption{A list of spectral lines used for the stratification calculations. The columns give the ion identification,
central wavelength, the excitation potential (in eV) of the lower level, oscillator strength ($\log\,{gf}$), 
the Stark damping constant (``appr'' marks lines for which the classical approximation was used), and the reference 
for oscillator strength. \label{strat-list}}
\begin{footnotesize}
\begin{center}
\begin{tabular}{lcrrrl|lcrrrl}
\noalign{\smallskip}
\hline
\hline
Ion &Wavelength &\ei\,(eV)  &$\log\,{gf}$&$\log\,\gamma_{\rm St}$ & Ref.&Ion &Wavelength &\ei\,(eV)  &$\log\,{gf}$&$\log\,\gamma_{\rm St}$& Ref.\\
\hline
\ion{Mg}{ii}&  3104.715&    8.864 &   -0.030&  -4.01  &KP   &\ion{Cr}{ii}&  5305.929&	10.760 &-0.170&  -5.348&RU  \\  
\ion{Mg}{ii}&  3104.721&    8.864 &   -1.330&  -4.01  &KP   &\ion{Cr}{ii}&  5308.425&	 4.071 &-2.060&  -6.639&RU  \\  
\ion{Mg}{ii}&  3104.805&    8.864 &   -0.190&  -4.01  &KP   &\ion{Cr}{i} &  5344.757&	 3.449 &-1.060&  -5.344&MFW \\  
\ion{Mg}{ii}&  4481.126&    8.864 &    0.740&  -4.70  &KP   &\ion{Cr}{i} &  5348.315&	 1.004 &-1.290&  -6.112&MFW \\  
\ion{Mg}{ii}&  4481.150&    8.864 &   -0.560&  -4.70  &KP   &\ion{Cr}{ii}&  5564.741&	10.893 & 0.510&  -5.364&RU  \\  
\ion{Mg}{ii}&  4481.325&    8.864 &    0.590&  -4.70  &KP   &\ion{Cr}{ii}&  5569.110&	10.872 & 0.860&  -5.359&RU  \\  
\ion{Mg}{i} &  4702.991&    4.346 &   -0.666&  -4.46  &LZ   &\ion{Cr}{ii}&  6050.242&	11.098 & 0.210&  -4.683&RU  \\    
\ion{Mg}{ii}&  4739.593&   11.569 &   -0.660&   appr  &KP   &\ion{Cr}{ii}&  6053.466&	4.745  &-2.220&  -6.633&RU  \\  
\ion{Mg}{ii}&  4739.709&   11.569 &   -0.820&   appr  &KP   &\ion{Cr}{ii}&  6138.721&	 6.484 &-2.150&  -6.728&RU  \\     
\ion{Mg}{i} &  5172.684&    2.712 &   -0.402&  -5.47  &AZ   &\ion{Cr}{ii}&  6147.154&	 4.756 &-2.890&  -6.656&RU  \\    
\ion{Mg}{i} &  5528.405&    4.346 &   -0.620&  -4.46  &LZ   &\ion{Cr}{ii}&  6336.263&	 4.073 &-3.760&  -6.638&RU  \\  
\ion{Mg}{i} &  6318.717&    5.108 &   -1.730&   appr  &KP   &	         & 	    &	       &      &	       &    \\  
\ion{Mg}{ii}&  7896.366&    9.999 &    0.650&  -4.54  &KP   &\ion{Fe}{ii}&  5018.440&	 2.891 &-1.340&  -6.585&RU  \\    
            &          &          &         &         &     &\ion{Fe}{ii}&  5018.669&	 6.138 &-4.010&  -6.537&RU   \\   
\ion{Si}{ii}&  4130.872&    9.839 &   -0.824&  -4.87  &BBCB &\ion{Fe}{i} &  5022.236&	 3.984 &-0.530&  -5.621&MFW \\    
\ion{Si}{ii}&  4130.894&   10.074 &    0.476&  -4.87  &BBCB &\ion{Fe}{ii}&  5022.420&	10.348 &-0.07 &  -5.367&RU  \\   
\ion{Si}{ii}&  4190.707&   13.492 &   -0.350&  -4.00  &AJPP &\ion{Fe}{ii}&  5022.583&	 5.571 &-4.180&  -6.622&RU  \\   
\ion{Si}{ii}&  4621.418&   12.256 &   -0.540&  -3.53  &NBS  &\ion{Fe}{i} &  5022.789&	 2.990 &-2.196&  -6.313&BWLW\\   
\ion{Si}{ii}&  4621.696&   12.256 &   -1.680&  -3.53  &NBS  &\ion{Fe}{ii}&  5022.792&	10.288 &-0.090&  -5.552&RU  \\   
\ion{Si}{ii}&  4621.722&   12.256 &   -0.380&  -3.53  &NBS  &\ion{Fe}{ii}&  5022.931&	 9.112 &-2.240&  -6.556&RU  \\   
\ion{Si}{ii}&  5055.984&   10.074 &    0.460&  -4.78  &BBCB &\ion{Fe}{i} &  5023.186&	 4.283 &-1.600&  -5.207&MFW \\   
\ion{Si}{ii}&  5056.317&   10.074 &   -0.490&  -4.78  &BBCB &\ion{Fe}{ii}&  5030.630&	10.288 & 0.431&  -5.891&RU  \\   
\ion{Si}{i} &  5701.104&    4.930 &   -2.000&  -4.41  &G    &\ion{Fe}{i }&  5030.778&	 3.237 &-2.830&  -6.277&BWLW\\
\ion{Si}{ii}&  5957.559&   10.067 &   -0.230&  -4.84  &BBCB &\ion{Fe}{i} &  5269.537&    0.859 &-1.321&  -6.300&BPS1\\   
\ion{Si}{ii}&  6371.350&    8.121 &   -0.003&  -5.04  &BBCB &\ion{Fe}{ii}&  5278.938&	 5.911 &-2.680&  -6.696&RU  \\   
            &          &          &         &         &     &\ion{Fe}{i} &  5281.790&	 3.038 &-0.834&  -5.489&BWLW\\    
\ion{Ca}{ii}&  3158.869&    3.123 &    0.241&  -5.54  &BWL  &\ion{Fe}{ii}&  5291.666&	10.480 & 0.540&  -5.468&RU  \\   
\ion{Ca}{ii}&  3933.655&    0.000 &    0.105&  -6.27  &BWL  &\ion{Fe}{ii}&  5303.395&	 8.185 &-1.530&  -5.822&RU  \\    			  
\ion{Ca}{i }&  4226.728&    0.000 &    0.265&  -6.03  &NBS  &\ion{Fe}{ii}&  5325.553&	 3.221 &-3.320&  -6.603&RU  \\   
\ion{Ca}{i }&  6162.173&    1.899 &   -0.167&  -5.32  &NBS  &\ion{Fe}{i} &  5326.142&	 3.573 &-2.071&  -6.209&BK  \\   
\ion{Ca}{i} &  6439.075&    2.526 &    0.390&  -6.07  &SR   &\ion{Fe}{i} &  5434.523&	 1.011 &-2.122&  -6.303&BPS1\\   
\ion{Ca}{ii}&  6456.875&    8.438 &    0.410&  -3.70  &TB   &\ion{Fe}{i} &  5560.211&    4.434 &-1.090&  -4.323&MRW \\   
\ion{Ca}{i} &  6462.567&    2.523 &    0.262&  -6.07  &SR   &\ion{Fe}{ii}&  5567.842&	 6.730 &-1.870&  -6.578&RU  \\    
            &          &          &         &         &     &\ion{Fe}{ii}&  5961.705&	10.678 & 0.670&  -4.950&RU  \\    
\ion{Cr}{ii}&  3180.693&    2.543 &   -0.319&  -5.30* &RU   &\ion{Fe}{i} &  6136.615&	 2.453 &-1.400&  -6.327&BPSS\\
\ion{Cr}{ii}&  3410.546&    4.781 &   -1.764&  -6.461 &RU   &\ion{Fe}{i} &  6137.691&	 2.588 &-1.403&  -6.112&BPS2\\   
\ion{Cr}{ii}&  3421.202&    2.421 &   -0.714&  -5.31* &RU   &\ion{Fe}{ii}&  6149.428&	 3.889 &-2.840&  -6.588&RU  \\    
\ion{Cr}{ii}&  3421.591&    4.294 &   -2.230&  -6.647 &RU   &\ion{Fe}{ii}&  6150.098&	 3.221 &-4.820&  -6.678&RU  \\     
\ion{Cr}{ii}&  3421.616&    4.316 &   -1.653&  -6.645 &RU   &\ion{Fe}{i} &  6335.330&	 2.198 &-2.177&  -6.195&BWLW\\    
\ion{Cr}{ii}&  3422.732&    2.455 &   -0.409&  -5.32* &RU   &\ion{Fe}{i} &  6336.824&	 3.686 &-0.856&  -5.467&BK  \\
\ion{Cr}{ii}&  5046.429&    8.227 &   -1.740&  -5.909 &RU   &		 &	    &          &      &	       &    \\ 
\ion{Cr}{i} &  5265.148&    3.428 &   -0.529&  -5.324 &K93  &\ion{Sr}{ii}&  3380.707&    2.940 & 0.199&   appr &W   \\  
\ion{Cr}{i} &  5296.691&    0.983 &   -1.400&  -6.120 &MFW  &\ion{Sr}{ii}&  3464.453&	 3.040 & 0.487&   appr &W   \\  
\ion{Cr}{i} &  5297.377&    2.900 &    0.167&  -4.307 &MFW  &\ion{Sr}{ii}&  3474.889&	 3.040 &-0.460&   appr &W   \\ 
\ion{Cr}{ii}&  5297.606&   10.754 &   -0.320&	 appr &RU   &\ion{Sr}{ii}&  4215.519&    0.000 &-0.145&  -5.50*&W   \\  
\ion{Cr}{i} &  5298.016&    2.900 &   -0.060&  -4.051 &MFW  &\ion{Sr}{ii}&  4305.443&    3.040 &-0.136&  -5.50*&W   \\  
\ion{Cr}{i} &  5298.272&    0.983 &   -1.150&  -6.117 &MFW  &\ion{Sr}{i} &  4607.327&    0.000 & 0.200&   appr &LW  \\  
\ion{Cr}{i} &  5298.494&    2.900 &   -0.350&  -3.749 &K93  &\ion{Sr}{i} &  4811.877&    1.847 & 0.190&   appr &GC  \\   
\ion{Cr}{ii}&  5305.865&    3.827 &   -2.160&  -6.599 &RU   &            &          &          &      &        &    \\
\hline                                
\end{tabular}
\end{center}
LZ -- Lincke \& Ziegenbein (\cite{LZ71}); KP -- Kurucz \& Peytremann (\cite{KP75}); AZ -- Andersen et al. (\cite{AZ67});
SG -- Schulz-Gulde (\cite{SG69}); G -- Garz (\cite{G73} -- corrected); BBCB -- Berry et al. (\cite{BBCB71}); AJPP -- Artru et al. (\cite{AJPP81}); 
NBS -- Wiese et al. (\cite{NBS69}); TB -- {\sc TOPBASE} (Seaton et al. \cite{TB92}); BWL -- Black et al. (\cite{BWL72});
SR -- Smith \& Raggett (\cite{SR81}); RU -- Raassen \& Uylings (\cite{RU98}); K93 -- Kurucz (\cite{K93}); MFW -- Martin et al. (\cite{MFW88});
MRW -- May et al. (\cite{MRW74}); BPS1 -- Blackwell et al. (\cite{BPS79}); BPSS -- Blackwell et al. (\cite{BPS82a}); 
BPS2 -- Blackwell et al. (\cite{BPS82b}); BWLW -- O'Brian et al. (\cite{OWL91}); BK -- Bard \& Kock (\cite{BK94}); 
W -- Warner (\cite{W68}); LW -- Lambert \& Warner (\cite{LW68}); GC -- Garcia \& Campos (\cite{GC88}).
\end{footnotesize}
\end{table*}

\begin{figure}[!t]
\figps{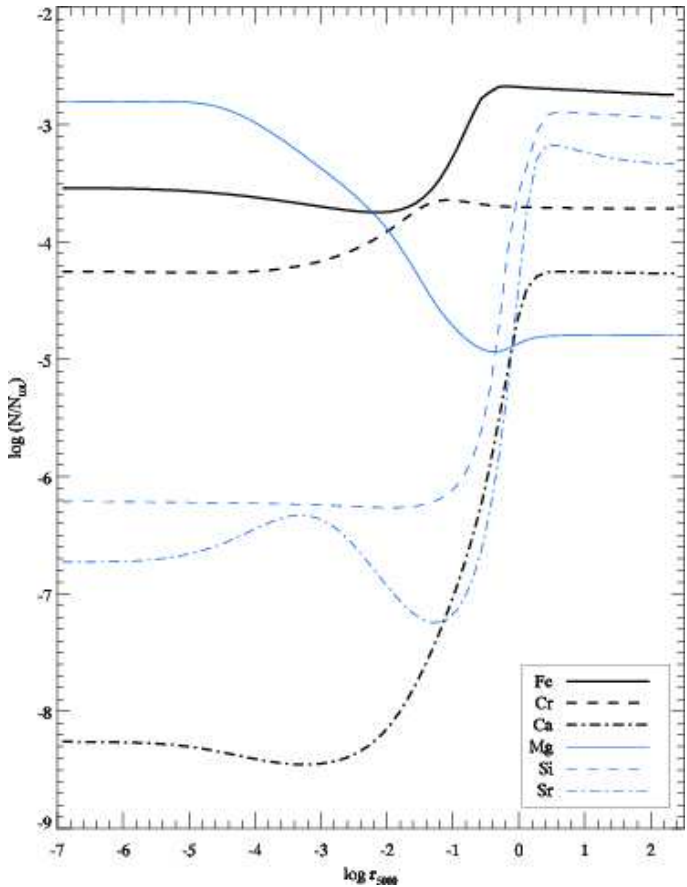}
\caption{Chemical stratification in the atmosphere of \hd.}
\label{fig5}
\end{figure}

\begin{figure*}[!t]
\figps{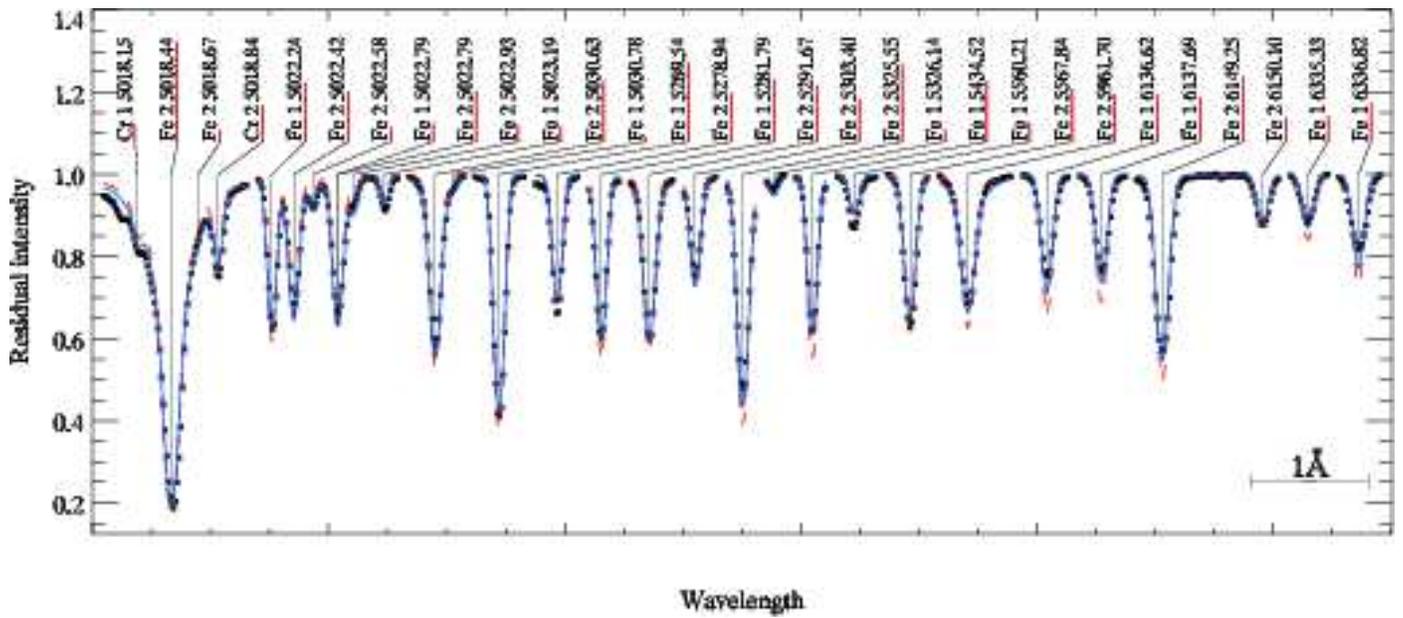}
\caption{Comparison of the observed Fe line profiles (\textit{symbols}) and calculations for
the stratified abundance distribution (\textit{full line}) and for the best-fit homogeneous abundance 
(\textit{dashed line}) of Fe.}
\label{fig6}
\end{figure*}

\begin{figure*}[!t]
\figps{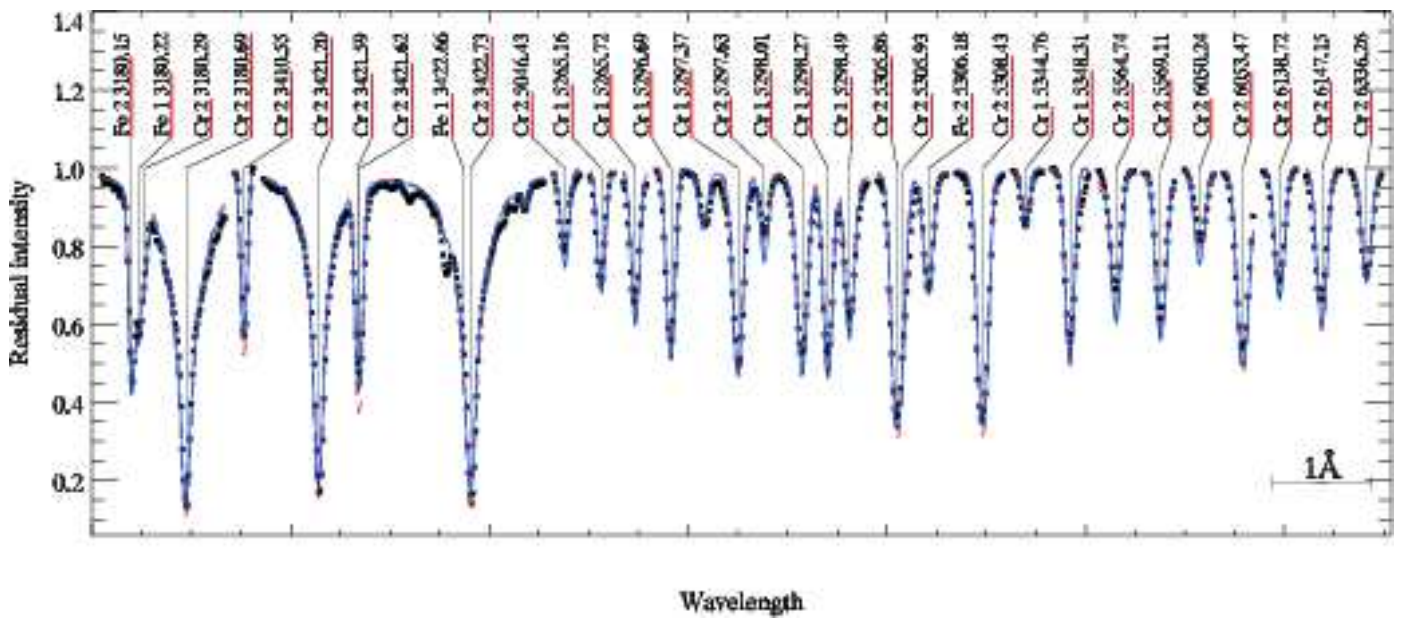}
\caption{Same as in Fig.~\ref{fig6} but for Cr.}
\label{fig7}
\end{figure*}

\begin{figure*}[!t]
\fifps{8.5cm}{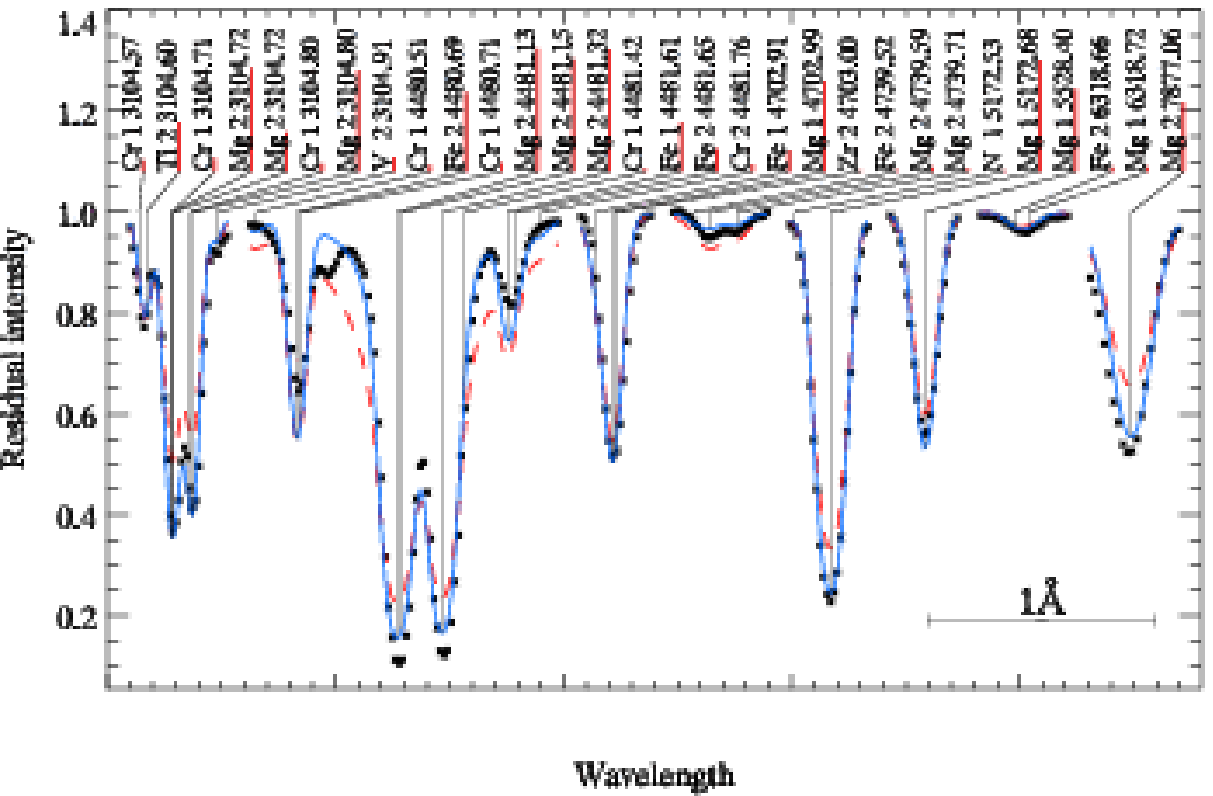}
\fifps{8.5cm}{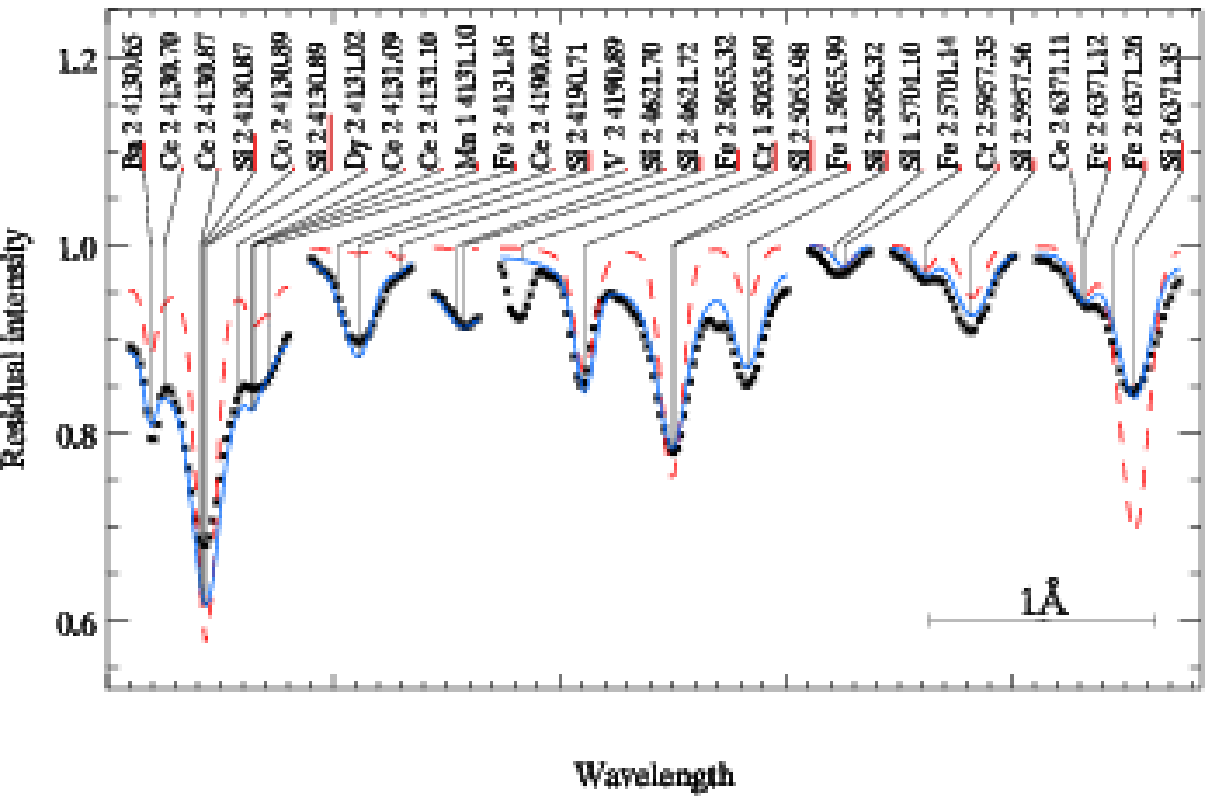}
\caption{Same as in Fig.~\ref{fig6} but for Mg (\textit{left panel}) and Si (\textit{right panel}).}
\label{fig8}
\end{figure*}

\begin{figure*}[!t]
\fifps{8.5cm}{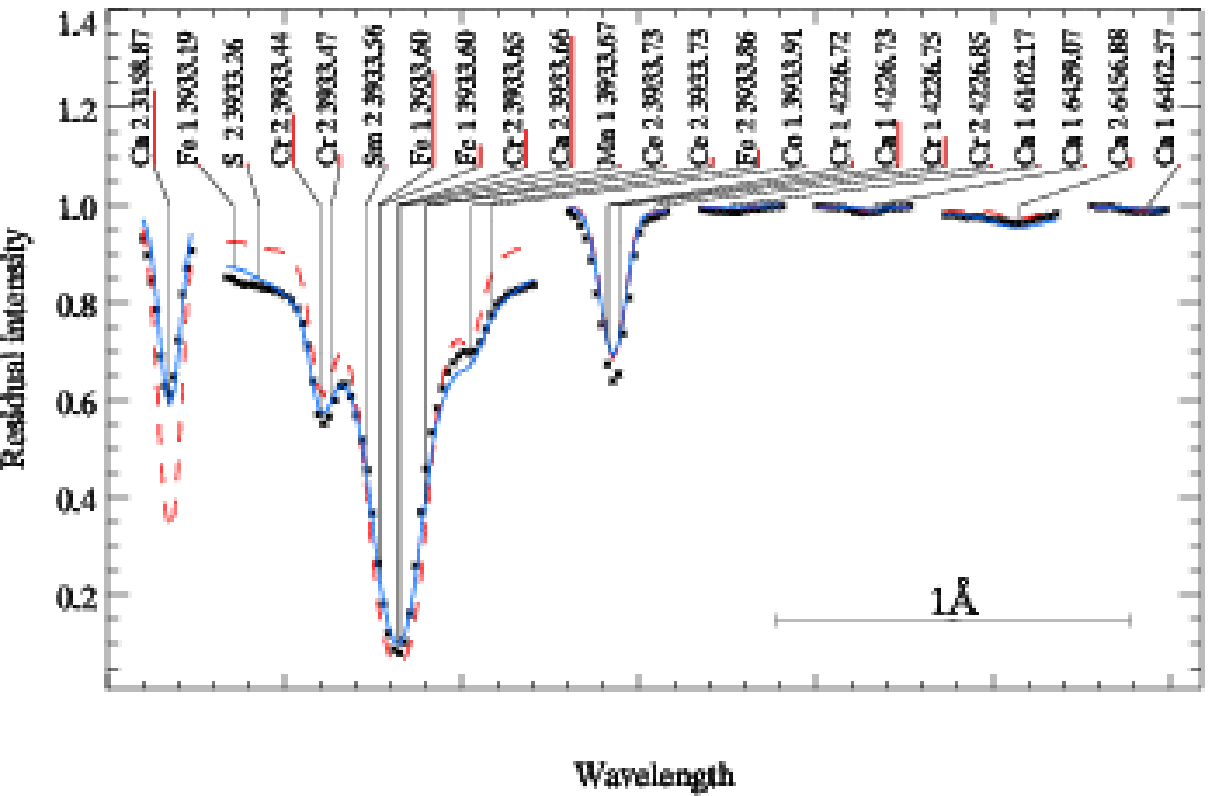}
\fifps{8.5cm}{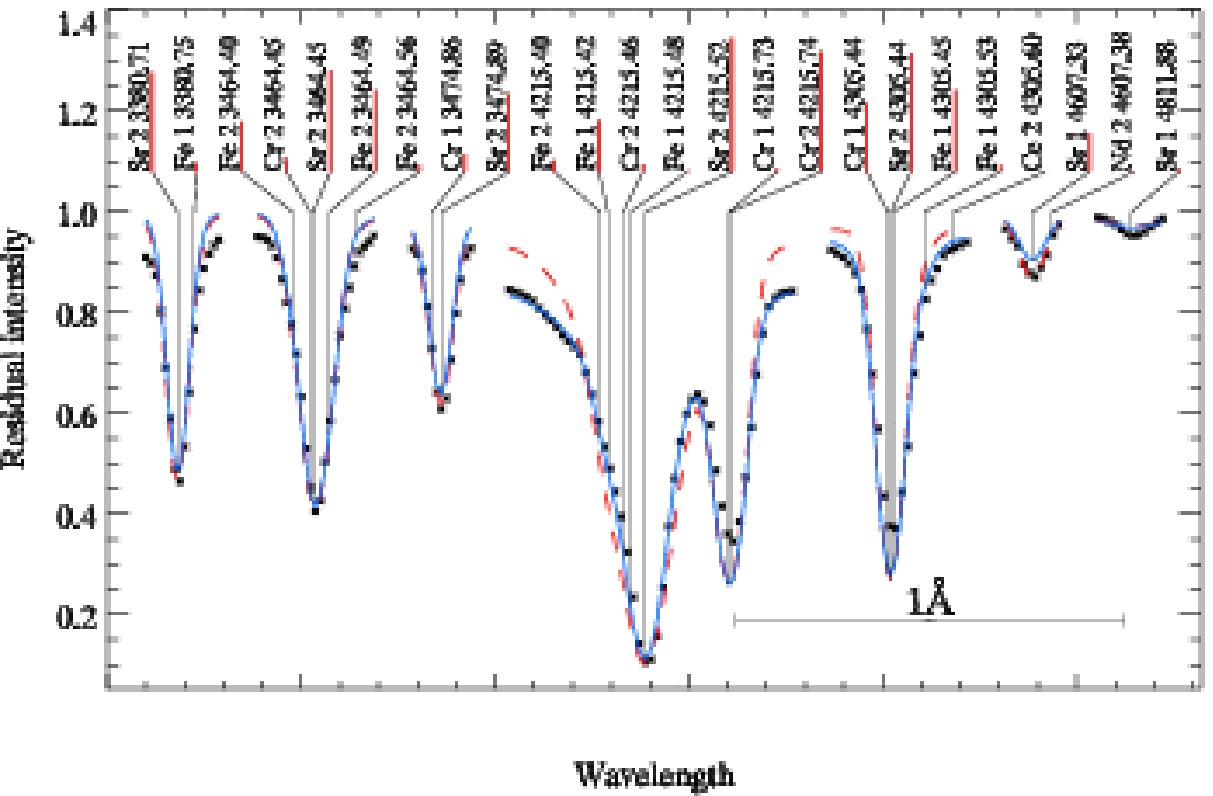}
\caption{Same as in Fig.~\ref{fig6} but for Ca (\textit{left panel}) and Sr (\textit{right panel}).}
\label{fig9}
\end{figure*}

The final depth-dependent abundances derived for the atmosphere of \hd\ using the \vip\ procedure are plotted
as a function of $\log\tau_{\rm 5000}$ in Fig.~\ref{fig5}. A comparison between observations and theoretical
spectra is presented in Figs.~\ref{fig6}--\ref{fig9}. Each of these plots shows observations and
the best-fit synthetic spectrum computed with the stratified abundance.
For comparison the best-fit calculation assuming a homogeneous chemical distribution is also
given.

\underline{\textit{Magnesium}} We have used 4 lines of \ion{Mg}{i}, one unblended \ion{Mg}{ii} line and 3
\ion{Mg}{ii} doublets to derive the vertical stratification of magnesium in \hd. This is the only element for
which abundance increases with height in the stellar atmosphere. The concentration of Mg is slightly below solar
in the deep layers and reaches a 1.7~dex overabundance above $\log\tau_{\rm 5000}\simeq-5.0$. The transition
zone extends over $\approx$\,4~dex in $\log\tau_{\rm 5000}$. Figure~\ref{fig8} shows that an inhomogeneous Mg
distribution substantially improves the fit to the \ion{Mg}{ii} lines, especially the feature at
$\lambda$~4481~\AA. The mean deviation between observations and spectrum synthesis -- a parameter that is used to quantify the
quality of the fit -- is lower by a factor of 2 for the model with a stratified Mg distribution.

\underline{\textit{Silicon}} This element shows outstanding stratification signatures in the spectra of \hd.
We have employed 7 \ion{Si}{ii} lines and one neutral Si transition for the reconstruction of the silicon
stratification in the atmosphere of \hd. We find that below $\log\tau_{\rm 5000}=0.0$ Si is overabundant by
1.5~dex with respect to the Sun. On the other hand, the upper atmospheric layers are characterized by a 1.7~dex
deficiency of Si. The transition zone is extremely narrow and is centred at $\log\tau_{\rm 5000}=-0.2$. A
dramatic improvement of the fit quality is evident (see Fig.~\ref{fig8}) for the theoretical spectrum which
takes the Si stratification into account. Such calculations allow us to reduce the mean deviation by as much
as a factor of 3 in comparison with the best-fit homogeneous model.

\underline{\textit{Calcium}} Ca is another element showing very strong vertical inhomogeneity in \hd. The Ca
distribution inferred for this star is similar to that of Si, but shows a larger change from the lower to
upper atmosphere. Ca is overabundant by 1.4~dex in the deep layers, whereas the deficiency with respect to 
solar composition reaches 2.5~dex in the shallow layers. Thus, the abundance of Ca in the atmosphere of \hd\
changes by almost 4 orders of magnitude. We obtain a 3.6 times lower mean deviation with the stratified Ca
model. These results are based on 4 \ion{Ca}{i} and 3 \ion{Ca}{ii} lines (see Fig.~\ref{fig9}). The resonance
\ion{Ca}{ii} K line is especially useful in the chemical inversion.

\underline{\textit{Chromium}} Many strong unblended spectral lines of neutral and singly ionized Cr are
identified in the spectra of \hd. For the \vip\ inversion we have employed 9 \ion{Cr}{i} and 17 \ion{Cr}{ii}
lines. Very high quality of the UV portion of the UVES spectra of \hd\ permitted analysis of the outstanding
\ion{Cr}{ii} 3180.29, 3421.20 and 3422.73~\AA\ lines (see Fig.~\ref{fig7}). We find weak evidence for the
presence of vertical stratification of Cr. This element is strongly enhanced over the whole line-forming
region in the photosphere of \hd. The Cr overabundance changes from 2.7~dex in the deeper layers to 2.2~dex in
the upper atmosphere. The stratified Cr distribution leads to a marginal ($\approx20$\% of the mean deviation)
improvement of the fit to the observed line profiles (Fig~\ref{fig7}).

\underline{\textit{Iron}} The vertical distribution of Fe was obtained from 11 \ion{Fe}{i} and 12
\ion{Fe}{ii} lines, including the strong line of \ion{Fe}{ii} at $\lambda$~5018.44~\AA\ (Fig.~\ref{fig6}).
Chemical inversion suggests a 1.9~dex enhancement in the deep layers and $\approx$\,1.0~dex overabundance in
the upper atmosphere, with the smooth transition at $\log\tau_{\rm 5000}=-1.0$. Comparison of the observed and
theoretical spectra computed  for the homogeneous and stratified Fe distribution is illustrated in
Fig.~\ref{fig6}. The case for the Fe stratification in \hd\ is somewhat stronger than for Cr: the inferred Fe
abundance step reaches 0.9~dex and the decrease of the mean deviation is more substantial ($\approx40$\%).

\underline{\textit{Strontium}} Our Sr abundance inversion relied on 5 \ion{Sr}{ii} lines and two weak
\ion{Sr}{i} features (Fig.~\ref{fig9}). The most important information about the vertical distribution of Sr in
\hd\ comes from the shapes of the profiles of strong \ion{Sr}{ii} lines. Accounting for the stratification of
this element noticeably improves the fit to the outer wings of the \ion{Sr}{ii} 4125.52 and 4305.44~\AA\
lines. A large overabundance of Sr is derived for the whole atmosphere of \hd. The Sr enhancement ranges from
5.9~dex deep in the atmosphere to 2.5~dex in the upper layers. Remarkably, the \vip\ inversion provides a
hint that the Sr vertical distribution is more complicated than the step-like profile found for other elements.

\section{Summary and discussion}
\label{discus}

Thanks to recent advances in the instrumentation at large telescopes, high-resolution, very high $S/N$
spectra have been obtained for a large number of Ap stars. This dramatic improvement of the data quality
stimulated a renewal of interest in the atmospheric properties of magnetic CP stars. Modern
investigations of the spectral line intensities and shapes show that many peculiar features in the Ap-star
spectra originate from an inhomogeneous vertical distribution of chemical elements in stellar atmospheres.
Thus, magnetic chemically peculiar stars represent the only type of non-degenerate stellar objects where
direct observations and diagnosis of the chemical diffusion signatures becomes possible. Observational
study of the chemical stratification in Ap stars is able to provide crucial and unique constraints for the
theoretical modelling of chemical diffusion in stellar atmospheres and also helps to understand the
relation between diffusion and hydrodynamic mixing processes.

Until now observational analyses of stratification in peculiar stars have been limited to fitting simple
parametrized vertical chemical profiles to a small number of diagnostic lines (Wade et al. \cite{WLRK03};
Ryabchikova et al. \cite{RPK02,RLK05}). Most often, chemical distributions were approximated with a step
function, whose shape is described by two abundance values and the position of the transition zone.

In the present study we have developed and applied the first assumption-free method of reconstruction of
the chemical stratification in stellar atmospheres. In our technique, the individual chemical composition is
derived for all atmospheric layers contributing to the line absorption. Uniqueness and stability of the
chemical inversion is achieved by applying the Tikhonov regularization function. This means that our code
finds the simplest elemental distribution sufficient to fit observations.

The vertical inversion procedure is successfully applied to the weakly magnetic, evolved Ap star \hd. We
have reconstructed vertical distributions of 6 elements. Magnesium is found to have an abundance close to the
solar one in the lower atmosphere and shows an increase of concentration with height. All other elements
have the opposite vertical distribution: high abundance deep in the photosphere and a lower abundance in the
upper layers. The transition region is narrow for Si, Ca and Sr, and is located close to $\log\tau_{\rm
5000}=0$. Fe and Cr are overabundant over the whole atmosphere of \hd, and the respective change of the
concentration of these elements is smaller than for other species and occurs between $\log\tau_{\rm 5000}=-1$
and $-2$.

Our investigation of the diffusion signatures in \hd\ confirms the validity of the step-like, parametrized
stratification models applied in previous studies of Ap stars. However, we find that for some species the
transition region can be rather extended (Mg), or the vertical distribution is more complex than a simple
one-step function (Sr). Thus, the applicability of the latter approximation is wide, but not universal. In
this context, our automatic regularized inversion approach appears to be more robust and should be
preferred whenever possible.

Stratification analysis of \hd\ is supplemented by a detailed chemical abundance study assuming
homogeneous vertical distribution of elements. We have measured the concentration of 43 ions of 32 chemical
elements. The outstanding characteristic of the chemical composition of \hd\ is a very large overabundance
of the iron-peak elements in its atmosphere, combined with a fairly small enhancement of heavy elements,
in particular REE. Strontium and, possibly, palladium are the only heavy elements in \hd\ showing a very large 
overabundance with respect to the solar chemical composition.

\hd\ is the second star with a detailed abundance analysis, including stratification effects, whose
position on the H-R diagram clearly shows that the star is close to finishing its main sequence life. The
first such star was HD~204411 (Ryabchikova et al. \cite{RLK05}). Both stars possess weak magnetic fields and  have a
small overabundance of the REEs compared with many other Ap stars in the 8500--9500~K effective temperature
region. The most remarkable abundance characteristic of both stars is an extremely high overabundance of Cr
and Fe, in particular in HD~133792. 

Cowley \& Henry (\cite{CH79}) have defined a group of 5 stars characterized by strong lines of the
Fe-peak elements and REE lines weaker than in ``normal'' Ap stars. HD~204411 is a member of this group.
The H-R diagram position of these objects indicates that all of them are close to the end of their main
sequence life (Kochukhov \& Bagnulo \cite{KB06}).

We were able to extract high-resolution spectra from the ELODIE  archive\footnote{{\tt
http://atlas.obs-hp.fr/elodie/}} for the three other stars of this group: 73 Dra (HD~196502), HD~8441, and
HD~216533. Photometric indicators suggest \te\,=\,8900--9200~K for these three objects. A preliminary 
determination of the Pr and Nd abundance based on the
strongest lines of the  second ions shows that in all three stars abundances of these rare-earth elements
are comparable to those found in \hd. In contrast, REE abundances in the young star HD~66318 of similar
effective temperature, much stronger magnetic field (Bagnulo et al. \cite{bagn03}) and the same large Cr and
Fe anomalies are  by an order of magnitude higher. For all stars these abundance estimates were based on 
the same lines of \ion{Pr}{iii} and \ion{Nd}{iii}. 

Evidently, on the basis of our detailed abundance analysis we may extend the definition of the group of stars
with weak REE lines as \textit{a group of evolved Ap stars with small to moderate REE anomalies, and very
large Cr and Fe overabundances}. All stars of this group are lie near maximum in plots of Cr and Fe abundance
versus \te\ (Ryabchikova et al. \cite{abun04}). Comparison between Cr and Fe abundances in evolved Ap
stars with weak magnetic fields and in the young star HD~66318 with an extremely large magnetic field shows
that neither evolutionary status nor magnetic field play an important role in creating the Fe-peak abundance
anomalies. On the other hand, the REE overabundance is clearly less pronounced in the group of evolved Ap
stars. Further detailed chemical and evolutionary analysis of Ap stars in the effective temperature range
8500--10000~K is required to understand the role of the magnetic field and/or evolution in creating and
maintaining REE atmospheric anomalies in magnetic peculiar stars.       

\begin{acknowledgements}
Resources provided by the electronic databases (VALD, SIMBAD, NASA's ADS)
are acknowledged. This work was supported by the grant from the Swedish \textit{Kungliga Fysiografiska
S\"allskapet}, by the Austrian \textit{Fonds zur F\"orderung der wissenschaftlichen Forschung}
(projects P17580N2, P17890) and BM:BWK (project COROT), and by the Presidium of RAS Program 
``Origin and Evolution of Stars and Galaxies''.
\end{acknowledgements}

\end{document}